\documentclass[
aps,%
12pt,%
final,%
notitlepage,%
oneside,%
onecolumn,%
nobibnotes,%
nofootinbib,%
superscriptaddress,%
noshowpacs,%
centertags]%
{revtex4}

\usepackage[english]{babel}
\usepackage{graphicx}
\usepackage{amsmath}
\usepackage{xcolor}
\usepackage{float}
\usepackage[caption = false]{subfig}
\begin{document}



\selectlanguage{english}
\title{Quasi-stationary tidal evolution
with arbitrarily misaligned  orbital and stellar angular momenta
with a preliminary numerical investigation in the  non-dissipative limit}
\author{\firstname{Pavel}~\surname{Ivanov}}
\email[]{pbi20@cam.ac.uk}
\affiliation{Astro Space Centre, P.N. Lebedev Physical Institute, 84/32
Profsoyuznaya Street, Moscow, 117997, Russia}
\author{\firstname{John}~\surname{Papaloizou}}
\email[]{J.C.B.Papaloizou@damtp.cam.ac.uk }
\affiliation{ DAMTP, Centre for Mathematical Sciences, University of
Cambridge, Wilberforce Road, Cambridge CB3 0WA}




\begin{abstract}

We review and extend the results of our 2021 paper  concerning the problem of tidal evolution of a binary system with a
rotating primary component with rotation axis  arbitrarily inclined with respect to the orbital plane. 
Only the contribution of quasi-stationary tides is discussed.
 Unlike  previous studies in this field we present evolution equations derived  'from first principles'. The governing equations contain two groups of terms.  

The first group of terms determines the evolution of orbital parameters and inclination angles
on a  long time scale determined by the rate of energy dissipation often described as  a 'viscous' time scale
though radiative damping may also be included. 
It can be shown that these terms are formally equivalent to corresponding expressions obtained by other 
authors after a map between the  variables adopted is established. 

The second group of terms is due to stellar rotation. These terms are present even when dissipation in the star is neglected. They may 
lead to conservative evolution of the angles specifying the orientation  of the stellar rotation axis and the orbital eccentricity { vector}
on a relatively short time scale. {  The corresponding 
evolution is also  linked to the rate of apsidal precession  of the binary orbit.}  Unlike in  our 2021 paper we consider
all potentially important sources of apsidal precession in an isolated binary, {  namely precession arising from the  tidal distortion
and rotation  of the primary as well as Einstein precession.
We solve these equations numerically for a small sample of input parameters,  leaving a complete analysis to an accompanying paper.   
Periodic changes to both the inclination  
of the rotational axis and its  precession  rate are found. 
In particular, for a particular binary parameters periodic flips between prograde and
retrograde rotation are possible.
 Also, when the inclination angle is allowed to vary,  libration of  the apsidal angle becomes possible.
  Furthermore, when the spin angular momentum is larger than the orbital
angular momentum there is a possibility of a significant periodic eccentricity changes similar those coming about from the well-known Kozai-Lidov effect. }
   
These phenomena could, in principle, be observed in systems with relatively large inclinations and eccentricities
such as e.g. those containing a compact object. In such systems both large inclinations and eccentricities could be generated as a result of a kick
applied to the compact object during a supernova explosion.

\end{abstract}

\maketitle

\section{Introduction}
Tidal interactions play an important  role 
in many close binary systems including those  containing a compact object, exoplanet systems, 
as well as  our own Solar system  ( see e.g.\citep{O2014} for a review).    
In this paper  we  begin by  reviewing  the main results obtained in our paper \citep{IP}, hereafter IP, where 
we considered tidal interactions, in the regime of quasi-static tides,  in
a binary system consisting of a primary component with spin angular momentum that is arbitrarily misaligned 
with the angular momentum of the orbit.  The companion is assumed to be compact and initially to have no internal degrees of freedom,
though  this was later relaxed to allow for  energy dissipation in its interior  that can contribute to the circularisation of the orbit.

This  configuration  setup  is the same as  that of  \citet{EKH},  hereafter EKH, who derived  the force and couple on a binary
orbit that arises from the dissipation associated with the equilibrium tide under the ad hoc assumption, that
the rate of dissipation of energy is  a positive definite function of the rate of change of the
primary quadrupole tensor,  viewed in a frame rotating with the star.  
Coriolis forces were neglected. 

Contrary to EKH, in \citep{IP} we calculate the response of the primary to tidal forcing from first principles in the low tidal forcing frequency limit,
starting from  the Navier-Stokes equations for a slightly perturbed  gravitating and slowly rotating 
gas sphere. We adopt  a quasi-static approximation, 
where in the leading approximation the stellar configuration is in hydrostatic equilibrium under the action of tidal forces. Motions associated with the equilibrium tide, dissipative processes and the effect of  the Coriolis force are included as next order corrections, up to  to first order in
the primary rotational frequency, { while neglecting the toroidal component of the response.} 
 
 Our approach removes any need for ad hoc assumptions about these phenomena such as connecting them
with the behaviour of the quadrupole tensor and provides a complete form for the response displacement
without the need for assumptions about unknown functions made by \citet{EKH}. We use the next order corrections mentioned above
to determine the effect of the tidal interaction on the orbit and angles characterising the orientation  of the primary's rotational axis. 
These terms lead to precession of the  primary rotation axis and orbital plane, which is added to that induced through
orbital torques and  stellar  centrifugal distortion,  in addition to  
 non-dissipative evolution of the angle
between  the orbital and spin angular momenta referred hereafter to as 'inclination angle'. 


It was shown by \cite{IP}, that the rate of the evolution of the inclination angle is determined by  apsidal precession rate.  In \cite{IP}  
the simplest situation where the apsidal precession is due only to the presence of quasi-static tides, (see e.g. \cite{St1939}) was considered.
 Here and in our accompanying 
paper \cite{IP1} we consider a more general setting  where  the apsidal precession is determined by all processes expected for an isolated binary system. 
Namely, we consider  relativistic Einstein precession,   precession arising from  tidal  distortion induced by quasi-static tides and apsidal precession arising from 
rotation of the primary, see e.g. \cite{BOC} and \cite{Sh}, \cite{PR} for a discussion of the latter term and its possible observational consequences.
Since the apsidal precession rate determined by rotational effects  depends on the inclination angle,  equations governing the evolution of the inclination 
and the longitude of periastron are coupled. Under certain conditions this coupling could lead to non-trivial effects, e.g. libration  of the apsidal line or 
a significant evolution of the inclination angle. Also, for systems with certain parameters, considerable evolution of the eccentricity results as a consequence of the 
conservation of total angular momentum of the system. In this paper we provide a preliminary numerical analysis of the corresponding evolution equations 
and {  identify certain potentially interesting regimes of  evolution. In an accompanying paper \cite{IP1} we provide an extensive analytic treatment of this
 problem. This enables one to find out which  effects, alone or in combination,  are important for  determining the apsidal precession rate, 
 and the  evolution of the angle of inclination between the orbital and spin angular momenta, for 
a given set of parameters of the system. }

The plan of this paper is as follows. In Section \ref{Basiceq} we give some  basic definitions and
define the three coordinate systems that are used to represent the dynamics of a binary with misaligned orbital  and primary spin
angular momenta. We introduce our set of equations for the determination of the orbital  and spin 
 angular momentum vectors in Section \ref{Jvecev}.  
{The form of the torque components acting on the star is then described in Section \ref{Inteval}.}
Expressions for the evolution of the semi-major axis and eccentricity are  given in Section \ref{andeev}. 
The potential contribution of non-dissipative terms  arising through the  effects of rotation,  including the Coriolis force, to the
orbital evolution are then discussed in Section \ref{NonrotT}.
In Section \ref{Discussion} we review and discuss our results.

\section{Basic definitions and equations}\label{Basiceq}
In this Section we  describe the basic model setup and coordinate systems as used in IP.
\subsection{Basic model}
For simplicity, we consider a binary system for which one of the components acts as  a point  mass, i.e. it has
no internal degrees of freedom referred to hereafter  as the companion \footnote{Note that  IP also discuss the 
possibility of an extended companion of low mass, whose spin is fully synchronised to the orbit.}.
The other component, referred to hereafter  as the primary possesses a distributed mass and a spin angular momentum which 
is unrestricted in orientation with respect  to the orbital angular momentum. Thus, both these angular momenta
are allowed to evolve with the resultant total angular momentum being conserved.

It is easy to show that the  evolution of the angular momentum vectors is  fully  determined
by  four simple governing equations  
following from the law of conservation of angular momentum  once
the torques
acting on the star are specified.  Calculation of the energy exchange with the orbit 
and the law of conservation of energy  then enables a complete description of the
evolution of the system once a prescription to determine the evolution of the orbital apsidal line
is prescribed. This is determined by the tidal interaction as well as the possible influence
of other orbiting bodies.  In the former case, to lowest order this is determined by the classical theory
of apsidal motion as indicated in IP.

\subsection{Coordinate system and notation}\label{Geometry}
We introduce three reference frames.
The first is a Cartesian coordinate system in a frame with origin at the centre of mass of the primary,  and for which the 
direction of the
conserved total angular momentum of the system,
${\bf J}$, defines the $Z''$ axis. The corresponding  $X''$ and $Y''$ axes are 
located in the orthogonal plane.

The second one is a Cartesian  frame such that  the orbital  angular momentum, ${\bf L}$, 
defines the direction of the $Z'$ axis. This is inclined to the total angular momentum vector, ${\bf J}$,  with an 
inclination $i$ which need not be constant as the orbital angular momentum is not conserved.

 The third
 $(X,Y,Z)$ coordinate system, described as the stellar frame,  is defined as in \cite{IP2011} with $z$-axis being directed along the direction of the stellar 
angular momentum vector, ${\bf S}$.
The azimuthal angle associated with  both ${\bf J}$ and ${\bf S}$  measured in the
$(X',Y',Z')$ system is $\pi/2 -\gamma.$ 
The $Y$ axis  lies in the  orbital  plane and  defines the line of nodes as viewed in  the $(X,Y)$
plane as in  \cite{IP2011}.
Note that the $ X',Y'$ and $Y$ axes are coplanar as are the  $Z, Z' $ and $Z''$ axes.
 For a Keplerian orbit with fixed orientation, the  line  of apsides can be chosen to coincide with the $X'$ axis.
In this case the angle between this line and the $X'$ axis, which we shall  more generally denote by $\varpi,$
will  simply be given by  
be $\varpi = 0 .$  Note that the angle between the apsidal line and the $Y$ axis, being the line of nodes can be taken 
quite generally to be  given by  $ \varpi +\gamma - \pi/2$ (with this choice it increases under positive rotation 
of the apsidal line in the orbit frame). 
  
\begin{figure}
\begin{center}
\vspace{1cm}
\includegraphics[width=14.0cm,height= 14.0cm,angle=0]{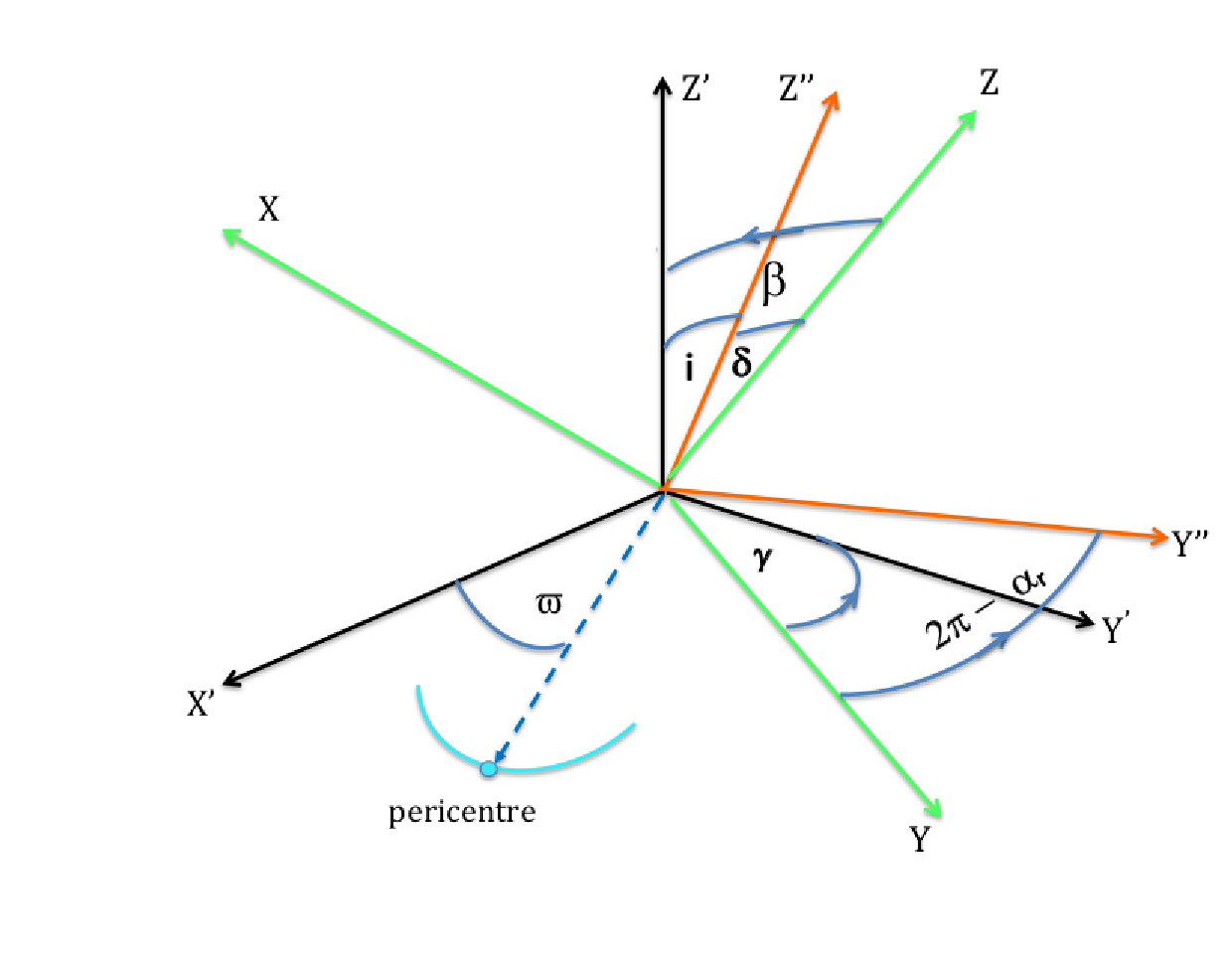}
\caption{Illustration of the $ (X,Y,Z)$  and   $(X',Y',Z')$ coordinate systems
together with the direction of the total angular momentum, which coincides with the $Z''$ axis
of  a coordinate system that is fixed in { the primary centred}  frame .
Note that the $ X', Y' $ and $Y$ axes are coplanar as are the  $Z, Z' $ and $Z''$ axes .
 The angle between the angular momentum vectors ${\bf L},$ directed along the $Z'$ axis 
and ${\bf S}$ directed along the $Z$ axis  is $\beta.$ { The angle between  ${\bf L}$ and the $Z''$ axis directed along ${\bf J}$ is $i$
and $\delta=\beta - i.$.   The angle between the $Y''$ axis and the $Y$ axis $2\pi -\alpha_r.$
The apsidal line, the location of pericentre and  an orbital arc in its neighbourhood 
are shown.}
}
\label{coordinates}
\end{center}
\end{figure}

The coordinate systems are illustrated in Fig. \ref{coordinates}.

\section{Evolution equations for the angular momentum vectors}\label{Jvecev}
 In this section we use the conservation of the total angular momentum to derive equations for the evolution
of the magnitudes  of the  orbital and spin angular momenta $L$ and $S$, and the angles $\beta$, $\delta=\beta-i$,  
and ${\bar \alpha }\equiv 2\pi -{\alpha_r},$ 
see Fig.  \ref{coordinates} for the definition of these angles.

From the definition of $\beta$ given above  it  follows that 
\begin{align}
&\cos {\beta}={({\bf L}\cdot {\bf S})\over LS}, 
\end{align}
where $L$ and $S$ are the magnitudes  of ${\bf L}$ and ${\bf S}$, and
\begin{align}
 &\cos {i}={({\bf J}\cdot {\bf L})\over JL},
 \end{align} 
 where $J$ is the magnitude  of
${\bf J=L+S}$.
 We  introduce  the torque ${\bf T}$ exerted on the star due to the tidal interaction. 
From the constancy  of the total  angular momentum  ${\bf J}={\bf L}+{\bf S}$ it follows that in the primary centred frame
\begin{equation}
{\bf T}=\dot {\bf S}=-\dot {\bf L}, 
\label{e0}   
\end{equation}
where a dot over a quantity,  
 here and subsequently, indicates the time derivative of that quantity.  
  
In addition, we also have
\hspace{2mm}  $2{\bf J}\cdot {\bf L} =J^2+L^2-S^2 $ \hspace{2mm}  and \hspace{2mm}  $2{\bf L}\cdot {\bf S} =J^2- L^2-S^2.$ 
and, accordingly the cosines of  $\beta$ and $i$ are given by,
\begin{equation}
\cos \beta ={J^2-L^2-S^2\over 2LS}\hspace{3mm}{\rm and} \hspace{3mm}\cos i = {J^2+L^2-S^2\over 2JL}.
\label{e41}
\end{equation}
From these relations we can also  express the sines of $\beta$, $i$ and $\delta$ in terms of $J$, $L$ and $S$, thus obtaining
\begin{equation}
\sin \beta =\frac{\sqrt{(J^2-(L-S)^2)((L+S)^2-J^2)}}{2LS},\hspace{2mm} {\rm and}
\hspace{2mm} \sin i = \frac{\sqrt{(S^2-(J-L)^2)((J+L)^2-S^2)}}{2JL}.           
\label{e39}
\end{equation}
In addition, consideration of the angular momentum components perpendicular to
{\bf J} and {\bf S} respectively gives
\begin{equation}
\sin \delta ={L\over S}\sin i = {L\over J} \sin \beta.
\label{e40}
\end{equation}

We now make use of rotation matrices, which
enable the transformation of the components of any vector from its representation in 
the stellar frame to its representation in the orbit frame,  and similarly its representation in  
the stellar frame to its representation in the
primary centred frame.
From these  transformations and their inverses (see IP for further discussion) 
it is straightforward to express the components of ${\bf L}$, ${\bf S}$
and the torque ${\bf T} = - \dot {\bf L}$  in the  primary centred  frame in terms of  $L$, $S$ and the components of the torque
in the  stellar frame \footnote{ We recall that by definition the components of ${\bf L}$ in the orbit frame are $(0,0,L)$
and the components of ${\bf S}$ in the stellar frame are $(0,0,S)$}.
								 
In this way the components of ${\bf L}$ in the  primary centred  frame,  $( L^{x^{''}},   L^{y^{''}}, L^{z^{''}}), $ are found to be  given by
\begin{equation}
L^{x^{''}}= L\cos {\bar \alpha }\sin i , \quad L^{y^{''}}=-L\sin {\bar \alpha} \sin i  \quad {\rm and}  \quad L^{z^{''}}=L\cos i.
\label{e42}
\end{equation}

\noindent Similarly, the components of ${\bf S}$ in the same  frame, $( S^{x^{''}},   S^{y^{''}}, S^{z^{''}}), $  are given by 
\begin{equation}
S^{x^{''}}= -S\cos{\bar  \alpha} \sin \delta , \hspace{2mm} S^{y^{''}}= S\sin {\bar \alpha }\sin \delta, \hspace{2mm} {\rm and}
\hspace{2mm}  S^{z^{''}}=S\cos \delta ,
\label{e43}
\end{equation}
and the components of ${\bf T}$ in this frame, $( T^{x^{''}},   T^{y^{''}}, T^{z^{''}}), $   are given by
\begin{equation}
T^{x^{''}}=\sin {\bar \alpha} T^y +\cos {\bar \alpha} T^{1}, \hspace{2mm} T^{y^{''}}=
\cos {\bar \alpha} T^y-\sin {\bar \alpha} T^{1}, \hspace{2mm} {\rm and} \hspace{2mm} T^{z^{''}}=\sin \delta T^x+ \cos\delta T^{z},
\label{e44}
\end{equation}
\begin{equation}
\hspace{-11.2cm} {\rm where}\hspace{2mm} T^{1}=\cos \delta T^x-\sin \delta T^z
\label{e44a}
\end{equation}
with the components of ${\bf T}$ in the stellar frame being given by
$(T^{x},T^{y},T^{z}).$

From equation (\ref{e0})  together with equations (\ref{e40}) - (\ref{e44a}) it is easy to obtain the following set of
equations 
\begin{equation}
{di\over dt}={1\over L}(-\cos \beta T^x +\sin \beta T^z), \quad {d L\over dt}=-\cos \beta T^z -\sin \beta T^x, \quad {d\over dt} {\bar \alpha }=
{T^y\over \sin i L}= {J\over LS}{T^y\over \sin \beta}, 
\label{e45}
\end{equation}
and
\begin{equation}
{d\over dt}\delta = -{T^x\over S}, \quad \dot S = T^z.
\label{e46}
\end{equation}
Note that we use (\ref{e40}) to obtain the last equality in (\ref{e45}). 
Also note that it is easy to check that $J^{i}=L^{i}+S^{i}$ are indeed  first integrals of
the set of equations  (\ref{e45}) and (\ref{e46}). 

{Using the first}  equation 
of (\ref{e45}) together with  (\ref{e46}) 
it is straightforward to obtain the evolution equation for angle $\beta$:
\begin{equation}
{d\beta \over dt}=-\left({\cos \beta \over L}+{1\over S}\right)T^x+{\sin \beta \over L}T^z.
\label{en1}    
\end{equation}
Together with the energy conservation law equations (\ref{e45}) and (\ref{e46}) form a complete set for
our model {once the torque components are specified.} 

Note that an evolution equation for the angle $\gamma$ is absent. This is because, physically,
only the angle ${\bar \alpha}=2\pi -\alpha_r$  appears in the specification of the orientation of the angular momentum
vectors in the { primary centred}  frame. In addition,  only
 the angle between the apsidal line and the  projection of the stellar  spin angular momentum vector onto 
the orbital plane, $\varpi+\gamma  $
matters for the determination of the orbital evolution. The evolution of this angle should be found from other considerations which are
not based on the law of angular momentum conservation. This evolution is determined, in general, from 
tidal interactions, stellar flattening, General Relativity and/or a presence of other perturbing
bodies, see below. 



\newpage
\section{The rate of change of the orbital elements}\label{RofCorb}

\noindent In order to specify how the orbital elements change we begin by providing expressions 
for the torque components acting in the stellar frame. 
\subsection{Expressions for the components of the torque acting on the star}\label{Inteval}

\noindent The torque components in the stellar  frame  to be used in equations (\ref{e45}) - (\ref{en1})  were derived in IP. 
The explicit expressions for them  are as follows:
\begin{align}
&\hspace{-0cm}{T}^{z} = T_*\left(
2\delta_1 \cos\beta
\phi_1
-\delta_2(1-e^2)^{3/2}
\left(\left (1+\cos^2\beta\right)
\phi_2- \sin^2\beta\cos2{\hat \varpi} \phi_3  \right)\right)\hspace{3mm}{\rm and}\hspace{3mm}
\label{t1} 
\end{align}
\begin{align}
&\hspace{0.5cm}T \equiv {T}^{x}-{\rm i}{T}^{y} = \nonumber\\
&\hspace{0.5cm} T_*\sin \beta ((2\delta_1-{\rm i}\delta_3)\phi_1
-( 1-e^2)^{3/2}(\delta_2-{\rm i}\sigma\delta_3)
\left((\phi_2+\phi_3\cos(2{\hat \varpi}))\cos\beta - {\rm i} \sin(2{\hat \varpi} )\phi_3\right)),
\label{t2} 
\end{align}
where $ \hat \varpi=\varpi + \gamma$. 
These expressions contain a number of dimensionless quantities we 
go on to define below.

\subsubsection{Dimensionless quantities appearing in the expressions for the torque components}\label{DIMT}
\noindent The three dimensionless quantities $\delta_1$,
$\delta_2$ and $\delta_3$  are determined by a strength of energy dissipation in the primary and its angular frequency of rotation, $\Omega_r$. Their
formal definitions are as follows
\begin{equation}
\delta_1=2\tilde \Gamma \tilde \Omega,\quad \delta_2 = \lambda\sigma \delta_1 
  ,\quad {\rm and }\quad \delta_3=-2\beta_*\lambda \sigma \tilde \Omega^2,
\label{e17}
\end{equation} 
where $\tilde \Gamma,$  defined as $\tilde \Gamma =\Gamma /\omega_{eq}$, is a characteristic rate of energy dissipation and
$\omega_{eq}$ is a characteristic dynamical time {defined in IP that is} associated with tidal disturbance of the primary. 

\noindent We expect that
 $\omega_{eq}~\sim~\Omega_{*}~\equiv~\sqrt{GM_*/ R_{*}^3}$, where $M_*$ and $R_*$ are the 
primary mass and radius, respectively. On the other hand it is expected that $\Gamma \ll \Omega_*$.
$\tilde \Omega =n_0/\omega_{eq}$, where $n_0$ is the orbital angular frequency \footnote{ In table 2 of IP ${\hat \Omega}$ erroneously appears instead of   ${\tilde \Omega}$ which should replace it.}.
\noindent  The dimensionless quantity  $\lambda = 2\beta_*/(2\beta_*+1)$, where $\beta_*$ 
is a dimensionless  parameter associated with the Coriolis force (see IP), it is expected to be smaller than, but order of unity.
 Similarly the dimensionless quantity  $\sigma = \Omega_r/(\lambda n_o) $,
is potentially  of order unity. 

\noindent Note  too that both $\delta_1$ and $\delta_2$ are proportional 
to the energy dissipation rate, 
while $\delta_3$ is determined by rotational effects.  Accordingly the terms proportional to $\delta_3$ lead to  non-dissipative 
evolution of the system.
\noindent The quantities $\phi_1,\phi_2,$ and $\phi_3$ are functions of the eccentricity, $e,$ given by 
\begin{align}
\phi_1=1+\frac{15}{2}e^2+ \frac{45}{8}e^4+\frac{5}{16}e^6,\label{phi1}
\end{align}
\begin{align}
\phi_2=1+3e^2+ \frac{3}{8}e^4 \hspace{3mm}{\rm and}\label{phi2}
\end{align}
\begin{align}
\phi_3=\frac{3}{2}e^2+ \frac{1}{4}e^4 .\label{phi3}
\end{align}

\subsubsection{General torque scaling}
All of the  torque components are proportional to a characteristic torque magnitude, $T_*,$ given by 
\begin{equation}
T_*= \frac{6\pi}{5}\left(\frac{ GM_pQ_{eq}}{ a^3(1-e^2)^3\omega_{eq}}\right)^2 = \frac{3k_2q^2}{1+q}
\left(\frac{R_*^5}{a^5}\right) \frac{M_*n_o^2a^2}{(1-e^2)^6},
\label{eq17n}
\end{equation} 
where $G$ is gravitational constant, $M_*$ and $M_p$ are respectively 
the masses of the primary star and compact component,  $q=M_{p}/M_{*}$ is the mass
ratio, $a$ is the orbital semi-major axis and  $Q_{eq}$ is the overlap integral  (see IP for its formal definition). 
The quantity $Q_{eq}$ has the dimensions of, 
$\sqrt{M_*}R_{*},$ and is expected to be smaller than one when expressed in this unit. The overlap integral can be  related to the apsidal motion
constant, $k_2,$ as can be seen by inspection of the  second equality in (\ref{eq17n}). 
\subsection{The rate of change of the orbital energy}
The rate of change of the orbital energy
is given by IP in the form
\begin{align}
&\frac{dE_{orb}}{dt} =\dot E_*
\left(\delta_2\phi_1 \cos\beta 
-\frac{\delta_1}{(1-e^2)^{3/2}}
\phi_4 
 \right),
\label{eng1} \\
&{\rm where} \hspace{10mm} \dot E_*=2n_o T_* \hspace{20mm}{\rm {and}}
\label{eqadd1}
\end{align}
\begin{align}
\phi_4=1+{31\over 2}e^2+ \frac{255}{8}e^4+\frac{185}{16}e^6  +\frac{25}{64}e^8. \hspace{3mm}
\label{phi}
\end{align}

\subsection{The evolution equations for the Euler angles and the angular momentum vectors}\label{Detorbspinev}
The rate of change of the absolute values of orbital and spin angular momentum vectors and the angles 
determining their orientation with respect to the { primary centred}  coordinate system follow from 
equations (\ref{e45}) and (\ref{e46}) after substitution of the components of the torque obtained from equations (\ref{t1}) and (\ref{t2}). 
Accordingly we obtain
\begin{align}
&\hspace{-3.3cm}{d i\over dt} = -(1-e^2)^{3/2}\sin \beta {T_*\over L}(\sigma \delta_3 \phi_3 \cos \beta \sin 2\hat \varpi 
+\delta_2 (\phi_2 - \phi_3 \cos 2 \hat \varpi)),    
\label{ev1}    
\end{align}
\begin{align}
&\hspace{-1.4cm}\frac{d \delta}{dt} = -{T_*\over S}\sin \beta\left( 2\delta_1 \phi_1
-(1-e^2)^{3/2}(\delta_2 \cos \beta (\phi_2 + \phi_3\cos 2\hat \varpi )+\sigma \delta_3 \phi_3 \sin 2\hat \varpi )\right),
\label{ev3}    
\end{align}
\begin{align}
&\hspace{-6mm}\frac{d  \alpha_r}{dt} = - \frac{d  \bar\alpha}{dt}= - {JT_*\over SL}\bigg( (\delta_3\phi_1-
(1-e^2)^{3/2}(\sigma \delta_3\cos \beta (\phi_2 + \phi_3 \cos 2\hat \varpi )+
\delta_2 \phi_3 \sin 2\hat \varpi ))\nonumber \\
&\hspace{-6mm} +{1\over 3}(1-e^2)^{9/2}{1+q\over q}\lambda^2 \sigma^2 \cos \beta \bigg)
\label{evn2}
\end{align}
and we note
that the rate of change of the  angle of inclination  between the spin and orbital angular momenta
is 
\begin{align}
&\hspace{-62mm}\frac{d\beta}{dt}  =\frac{d i }{dt}+\frac{d \delta}{dt} .\hspace{60mm}. 
\end{align}
The rate of change of the magnitudes of the orbital and spin angular momenta are respectively given by
\begin{align}
&\hspace{-15mm}\frac{d L}{dt} = -2 T_* (\delta_1 \phi_1-(1-e^2)^{3/2}\delta_2\phi_2 \cos \beta)-(1-e^2)^{3/2}T_* \sigma \delta_3 \phi_3 \sin^2 \beta
\sin 2\hat \varpi,
\label{ev2}    
\end{align}
and 
\begin{align}
&\hspace{-5.8cm} \frac{d S}{dt} = T^z,\hspace{7.2cm}
\label{ev4}
\end{align}
where $T^z$ is given by equation (\ref{t1}). 

Note that in addition to the 
 expression for the 
torque component $T^y$  in equation (\ref{evn2}) given by equation(\ref{t2}) (see equation(\ref{e45}))  we have  added
a contribution $T^y_{SF}$ arising
from the effect of stellar flattening due to rotation. 
This produces the last term in  equation (\ref{evn2}) 
which  has  the factor $(1+q)/ q$, where $q=M_p/M_*$ is the mass ratio.
 While our 'standard' torque component $T^y$ is proportional to the stellar rotational frequency $\Omega_r$, 
 $T^y_{SF}$ is proportional to the square of $\Omega_r,$ hence the factor $\sigma^2$ 
 in this  term.

We recall that $J$ is the 
conserved total angular momentum of the system, while $i$ and $\delta$ are,  respectively, the angles of inclination
between this and the orbital and spin angular momenta. 
In addition  the quantities $\phi_i$ are given by
equations (\ref{phi1})-(\ref{phi3}) and (\ref{phi}) with $T_*$ and ${\dot E^*}$ being given by equations (\ref{eq17n}) and  (\ref{eqadd1}).

\subsection{Evolution of the semi-major axis and eccentricity}\label{andeev}
\noindent For Keplerian orbits the relationship between the rate of change of the 
semi-major axis and the rate of 
change of orbital energy is given by 
\begin{equation}
\frac{d a}{dt} = \frac{2a^2}{GM_pM_*}\frac{dE_{orb}}{dt}=  
\frac{2a^2}{GM_pM_*}\dot E_*
\left(\delta_2\phi_1 \cos\beta  
-\frac{\delta_1\phi_4}{(1-e^2)^{3/2}}\right),  \label{jp18e1}
\end{equation}
where we have used the expression for $dE_{orb}/dt$  given by equation (\ref{eng1}).
The  rate of change of the
orbital eccentricity is given in terms of the  rates of
change of orbital angular momentum and  energy  by
\begin{equation}
\frac{d e}{dt} =\frac{a(1-e^2)}{GM_pM_*e} \left(\frac{dE_{orb}}{dt}-
\frac{d L}{dt}\frac{\sqrt{G(M_p+M_*)}}{a^{3/2}\sqrt{1-e^2}}\right)\label{jp1121}
\end{equation}
Substituting (\ref{eng1}) and (\ref{ev2}) in (\ref{jp1121})
we obtain
\begin{align}
&\dot e =-{3a e(1-e^2)^{-1/2}\dot E_*\over GM_pM_*}(3\delta_1 \phi_5-{11\over 6}\delta_2\phi_6(1-e^2)^{3/2}\cos \beta)-\nonumber\\
&{3\over 4}{a e (1+e^2/ 6)(1-e^2)^2\dot E_*\over
GM_pM_*}\sigma \delta_3 \sin^2 \beta \sin 2\hat\varpi ,
\label{ev5}    
\end{align}
where we  use  equation (\ref{phi3}) to obtain $\phi_3$ and
\begin{equation}
\phi_5=(\phi_4-(1-e^2)\phi_1)/(9e^2)=
1+{15\over 4}e^2+{15\over 8}e^4 +{5\over 64}e^6
\label{ev6}    
\end{equation}
and
\begin{equation}
\hspace{-6mm} \phi_6=\frac{2(\phi_1-(1-e^2)\phi_2)}{11e^2}=1+{3\over 2} e^2
+{1\over 8}e^4.
\label{ev7}    
\end{equation}

\section{The conservative orbital evolution resulting from  rotational effects}\label{NonrotT}
As mentioned above in Section \ref{DIMT}  our evolution equations contain two types of terms - those proportional to $\delta_1$ and $\delta_2$, and,
accordingly, to the energy dissipation rate  characterised by, $\Gamma$, and those proportional to $\delta_3$, which are associated with
conservative rotational effects.
The former group of terms may be shown to lead to an  equivalent 
description to that discussed in EKH after an appropriate mapping  between the two formalisms is made (see IP). 
However, note that,
 our formalism has the advantage that it allows one to relate directly the quantities governing the system evolution
 driven by  the energy dissipation rate
 to well defined properties of the primary star. 
 The latter group of terms leads to qualitatively new effects. We review them below.

Considering the terms in the tidal response  that are $\propto \delta_3$
we neglect dissipative effects in the primary.  formally setting $\delta_1= \delta_2=0$.
 The tidal interaction
then conserves orbital energy and consists of an interaction between the spin and orbital angular momenta,
characteristically leading to changes in their mutual inclination, accompanied by their precession around 
 the total angular momentum vector.
It is important to
stress that this approximation may be adequate for sufficiently short time intervals, since the
dimensionless parameters
$\tilde \Gamma^{-1} $ and  $\tilde\Gamma_{p}^{-1}$ which  determine  the timescale of  evolution 
due to the presence of non conservative effects  are  expected to be 
relatively quite large.

In addition to the  conservation of the orbital
energy, it follows  from the fact that the $Z$ component of the  torque $T^z=0$ when $\delta_{1,2}=0$  
that the magnitude of the  rotational angular momentum,
$S$, is an integral of the motion. Also, from the second  equation in the set of
(\ref{e45})  together with equation (\ref{en1}) it follows that there is additional integral
of motion
\begin{equation}
I={L^2\over 2S}+L\cos \beta,   
\label{ev11}    
\end{equation}
which is valid for any form of $T^x$.\footnote{Note a misprint in IP, where there should be $-$ instead of 
$+$ in (\ref{ev11}).}

Since eccentricity $e$ depends only on $L$ when the orbital
energy, and, accordingly, semi-major axis $a$ are fixed, from equations (\ref{en1}) and (\ref{t2})   it follows
that the evolution equation for ${\dot \beta}$ is a function 
of $\beta$, $e$ and $\hat \varpi$ when $a$ and $S$ are fixed. From equations (\ref{en1}) and (\ref{t2}), where
we set we set $\delta_2=0$ and $\delta_3=-2\beta_*\lambda \sigma
{\tilde \Omega}^2$ and remember that $\tilde \Omega =n_0/\omega_{eq}$ it is  seen that it has 
the form
\begin{align}
\dot \beta=\left( \frac{1}{S}+ \frac{\cos\beta}{L} \right) T_*   \frac{3(2\beta_*+1) e^2(1-e^2)^{3/2}}{2\tilde \omega^2_{eq}}\left(1+\frac{e^2}{6}\right)
\left( \frac{\Omega_r}{\Omega_{*}} \right)^2   \sin\beta\sin{2\hat \varpi}, \label{betaeq}
\end {align}
where $\Omega_{*}=\sqrt{GM_*/R_*^3}$ is a typical 'stellar' frequency and $\tilde \omega_{eq}=\omega_{eq}/\Omega_{*}\sim 
1$. 

The evolution equation for $e$ follows from (\ref{ev5}) 
\begin{align}
&\dot e =-{3\over 4}{(2\beta_*+1)\over \tilde \omega^2_{eq}}{a e (1+e^2/ 6)(1-e^2)^2\dot E_*\over
GM_pM_*}\left( \frac{\Omega_r}{\Omega_{*}} \right)^2 \sin^2 \beta \sin 2\hat\varpi,
\label{eccentricity}    
\end{align}
where we remember that $\lambda ={2\beta_* /( 2\beta_*+1)}$.  
Note that the same equation can be also obtained by differentiating (\ref{ev11}) in time and substituting (\ref{betaeq}) into
the result.

In \cite{IP} it was  assumed that the rate of change of ${\hat\varpi}$ { is determined by classical tidal distortion  and is given by}
\begin{align}
\frac{d{\hat\varpi}}{dt} = \frac{d{\varpi}}{dt}\equiv \frac{d{\varpi_T}}{dt} =15 k_2n_0\frac{M_pR_*^5}{M_*(a(1-e^2))^5}\phi_6,\label{Apse}
\end{align} 
 ( see e.g. \cite{St1939}), where $\phi_6$ is given by eq. (\ref{ev7}).
and (\ref{ev2}) still applies. In \cite{IP1} and in this paper we consider all potentially important effects causing apsidal precession in an isolated binary star with one point-like component:
\begin{align}
\frac{d{\varpi}}{dt}=\dot \varpi_{T} +\dot \varpi_{E} + \dot \varpi_{R} + \dot \varpi_{NI},
\label{prec}
\end{align}     
where ${d \varpi_{T} / dt}$ is given by (\ref{Apse}), 
\begin{align}
{d\varpi_{E} \over dt}={3GM_*(1+q)\over c^2a(1-e^2)}n_{0}, \label{Ein}
\end{align}
is the standard expression for the Einstein relativistic apsidal precession, $c$ is speed of light, and 
\begin{equation}
\dot \varpi_{R}=-{(3\cos^2 \beta -1)\over 2}\Omega_{Q}, \quad \dot \varpi_{NI}=-{J\over S}\cos \beta \cos i \Omega_{Q}\equiv -{ (L+S\cos\beta)\over S}\cos \beta  \Omega_{Q},
\label{a15}
\end{equation}
with
\begin{equation}
\Omega_Q=-{k_2(1+q)\over (1-e^2)^2}{\left({R_*\over  a}\right)}^{5}{\left({\Omega_r\over n_0}\right)}^2n_{0},   
\label{a3} 
\end{equation}
being determined by rotational flattening of the primary star. 
{ Derivation of the  expressions (\ref{a15}) are given by \cite{IP1} making use of} a general expression provided by \cite{BOC}. Physically,  $\dot \varpi_{R}$ is determined 
by corrections to Keplerian gravitational potential of the primary star due to its rotational flattening, while $\dot \varpi_{NI}$ is due to evolution 
of the orbital frame, which leads to non-inertial effects causing additional apsidal precession. 

Remembering that  $a$ and $\Omega_r$ are constant in case of non-dissipative evolution, using the standard expression 
for the orbital angular momentum $L$ and expressing the stellar angular momentum in terms of $\Omega_r$ 
as $S=I\Omega_r$, where $I$ is stellar moment of inertial, it is easy to see that 
 equations (\ref{betaeq}), (\ref{eccentricity}) and (\ref{prec}) form a complete set. 

\section{Some examples of numerical solutions of the  dynamical evolution equations}\label{VI}

We solve equations (\ref{betaeq}), (\ref{eccentricity}) and (\ref{prec}) by a standard Runge-Kutta procedure of forth order.
We use the 'slow' time variable $\tau=t/t_*$, where
\begin{equation}
t_*={{\tilde a}^{13/2}\Omega_*^{-1}\over 15 k_2 q\sqrt{(1+q)}}, 
\label{T} 
\end{equation}
where we have introduced a dimensionless semi-major axis $\tilde a=a/R_*$. 
From equation (\ref{Apse}) it is seen that $t_{*}$ is
a characteristic timescale of tidal apsidal precession at small values of eccentricity.

 We assume that  the dimensionless quantities $\tilde I=I/(0.1M_*R_*^2)$,
$\alpha_{E}=\left({M_*/ M_{\odot}}\right)\left({k_2/10^{-2}}\right)^{-1}\left({R_*/ R_{\odot}}\right)^{-1}$ and    
$\gamma_{*}=(2\beta_{*}+1)/(2 \omega_{*}^2)$ are all equal to unity. We consider three different values of mass ratio $q=M_p/M_*$, 
$q=10^{-3}$, $q=1$ and $q=10^{3}.$ In the first two cases the  dimensionless semi-major axis is set to be $\tilde a=5$, while in 
the large mass ratio case it is set to be $\tilde a=50$.  Calculations are performed for a range of initial inclination
angles, $\beta_0.$
The dimensionless rotation frequencies adopted are  $\tilde \Omega_r=\Omega_r/n_0=0.1$, $1$ and $3.$  The initial eccentricities were 
$e_0=0.5$ and $0.7$ while the  initial value of the apsidal angle $\varpi_0={\pi/ 4}$ for all runs.

All runs start at $\tau=0$ and end when $\tau=\tau_{end}$,
where $\tau_{end}={6\pi \over t_{*}\dot \varpi_{max}}$ and $\dot \varpi_{max}=\sqrt{{\dot \varpi_{T}}^2 +{\dot \varpi_{E}}^2 + {\dot \varpi_{R}}^2 + {\dot \varpi_{NI}}^2}$. Since we are going to present only results of a rather limited number of runs { that extend over a limited time span,} 
 and the parameter space of the problem is quite large it is clear that the results presented here should be considered as being  {only preliminary and  indicative}. 

When choosing values of $\tilde a$ and $\tilde \Omega_{r}$ it is important to remember that those values are constrained by physical conditions, ( see also \cite{IP1}). At first,
it is clear that the orbit periastron, $r_{p}=(1-e)a$, should be larger than the stellar radius $R_{*}$. Secondly,
periastron cannot be smaller than than tidal disruption 
radius $r_{T}={({M_1/ M_*})}^{1/3}R_*=q^{1/3}R_{*}$. In general, we have
\begin{equation}
\tilde a > \tilde a_{min}={\max (1, q^{1/3}) \over (1-e)}.
\label{lim1} 
\end{equation}
Also, the rotational frequency $\Omega_r$ should be smaller by some factor than $\Omega_{*}$ for our theory to 
be valid. As in \cite{IP1}  we assume that $\Omega_{r} < 0.5\Omega_{*}$ (see e.g. \cite{IP2007a} for a discussion), thus obtaining
\begin{equation}
\tilde \Omega_r < \tilde \Omega_{max}={{\tilde a}^{3/2}\over 2\sqrt{1+q}}.
\label{lim2} 
\end{equation}

In \cite{IP1} it is shown that when $\tilde \Omega_r$ is large and $\beta_0$ is chosen in such a way when 
cancellation between different terms in (\ref{prec}) does not occur,  a typical variation of $\beta$, $\Delta \beta$,
is expected to be
\begin{equation}
\Delta \beta \sim {9}q
{e^2(1+{e^2/ 6})\over (1-e^{2})^{3}} {\tilde a}^{-3}\sin \beta_0. 
\label{dbet} 
\end{equation} 
In what follows we  compare (\ref{dbet}) with the results of our numerical runs.   {  It was also shown in \cite{IP} 
that 
 cancellation between terms on the right hand side of (\ref{prec})  is  expected close to 'critical' values of $\beta_0$, $\beta_{crit}$, defined by the condition that $\cos^{2}(\beta_{crit})={1/ 5}$ and on some 'critical' curves $\tilde a=a_{crit}(\tilde \Omega_r, \beta_0)$ defined by the condition that the right hand side of (\ref{prec}) is zero on the curve. }
From the condition  $\cos^{2}(\beta_{crit})={1/ 5}$ we have 
$\beta_{crit,\pm}\approx 1.107$, $2.03$ corresponding  
to prograde and retrograde rotation, respectively. Note that  critical curves  exist only when $\beta > \beta_{crit,+}$. When a solution crosses the critical curve in the course of its evolution its behaviour changes drastically. In this case we expect large variations of $\beta$ and oscillatory (librating) behaviour of the apsidal angle. We demonstrate explicit examples of such solutions below.       

In general, it is expected that the evolution of our system is periodic, with a period smaller than $\tau_{end}$. During this evolution the angle $\beta $ changes periodically, while $\hat \varpi$ could either secularly  increase its value (circulate) or have an oscillatory behaviour (librate). 
In order to characterise the behaviour of our dynamical system over the run time
we introduce $\Delta \varpi =\hat \varpi - \varpi_{0}$ and $\Delta \beta = \beta - \beta_{0}$. Since it is expected that 
$\Delta \beta $ may change its sign over the evolution we evaluate its maximum value over the time interval  $(0,\tau_{end})$, and plot it as a function of the system parameters and initial values of the dynamical variables.  We also plot the final value of $\Delta \varpi$.

\subsection{The case $q=10^{-3}$}    

\begin{figure}
\begin{center}
\vspace{1cm}
\includegraphics[width=14.0cm,height= 14.0cm,angle=0]{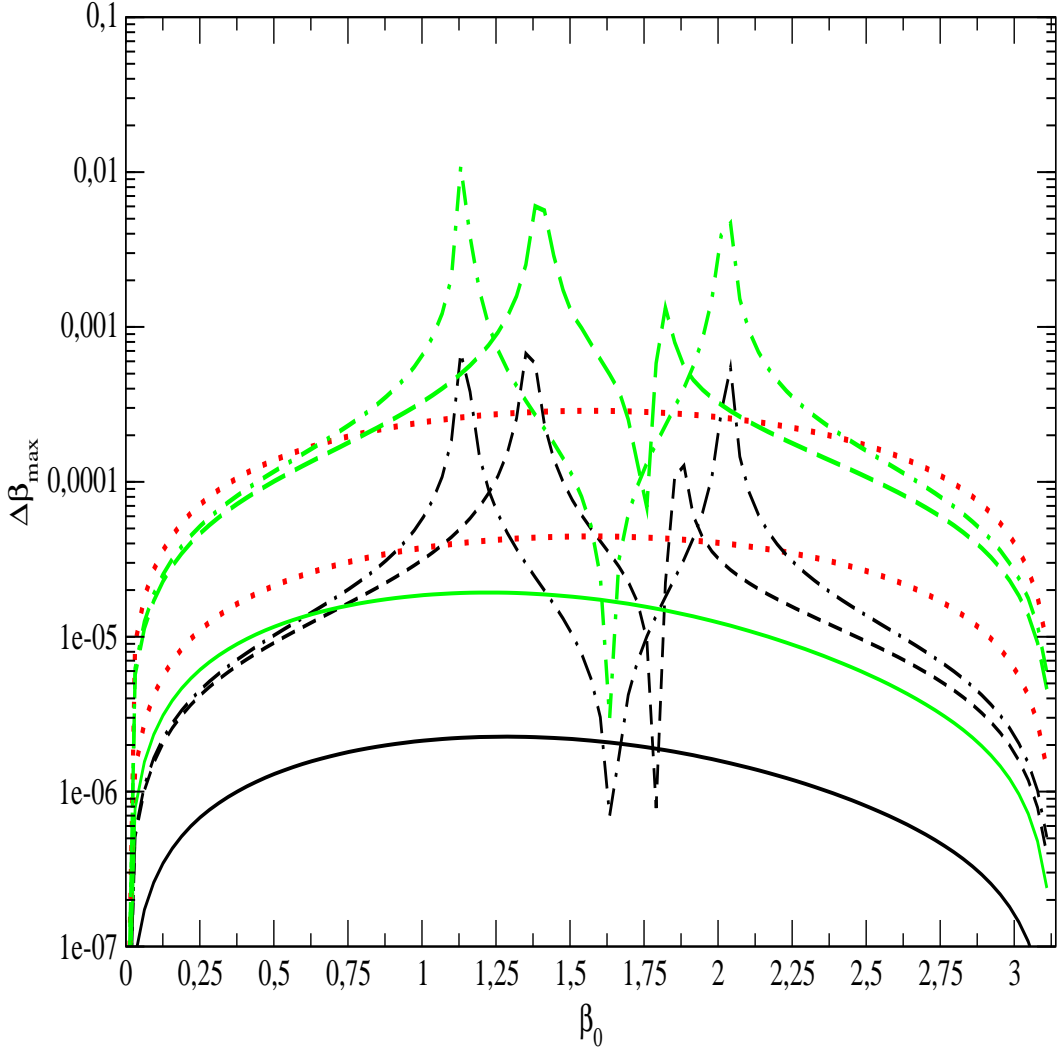}
\caption{The maximum value of $\Delta \beta$,   $\Delta \beta_{max}$, during a run  is shown as a function of the initial inclination angle $\beta_0$  for $q=10^{-3}$. 
Solid, dashed,  and dot dashed curves are for $\tilde \Omega_{r}=0.1$, $1$ and $3$, respectively. Dotted curves
represent the { maximum value of the}  expression (\ref{dbet}). Curves of the same type 
with larger (smaller) values of their arguments for a given $\beta_{0}$ are calculated for $e_{0}=0.7$ ($0.5$).}
\label{fig1}
\end{center}
\end{figure}

\begin{figure}
\begin{center}
\vspace{1cm}
\includegraphics[width=14.0cm,height= 14.0cm,angle=0]{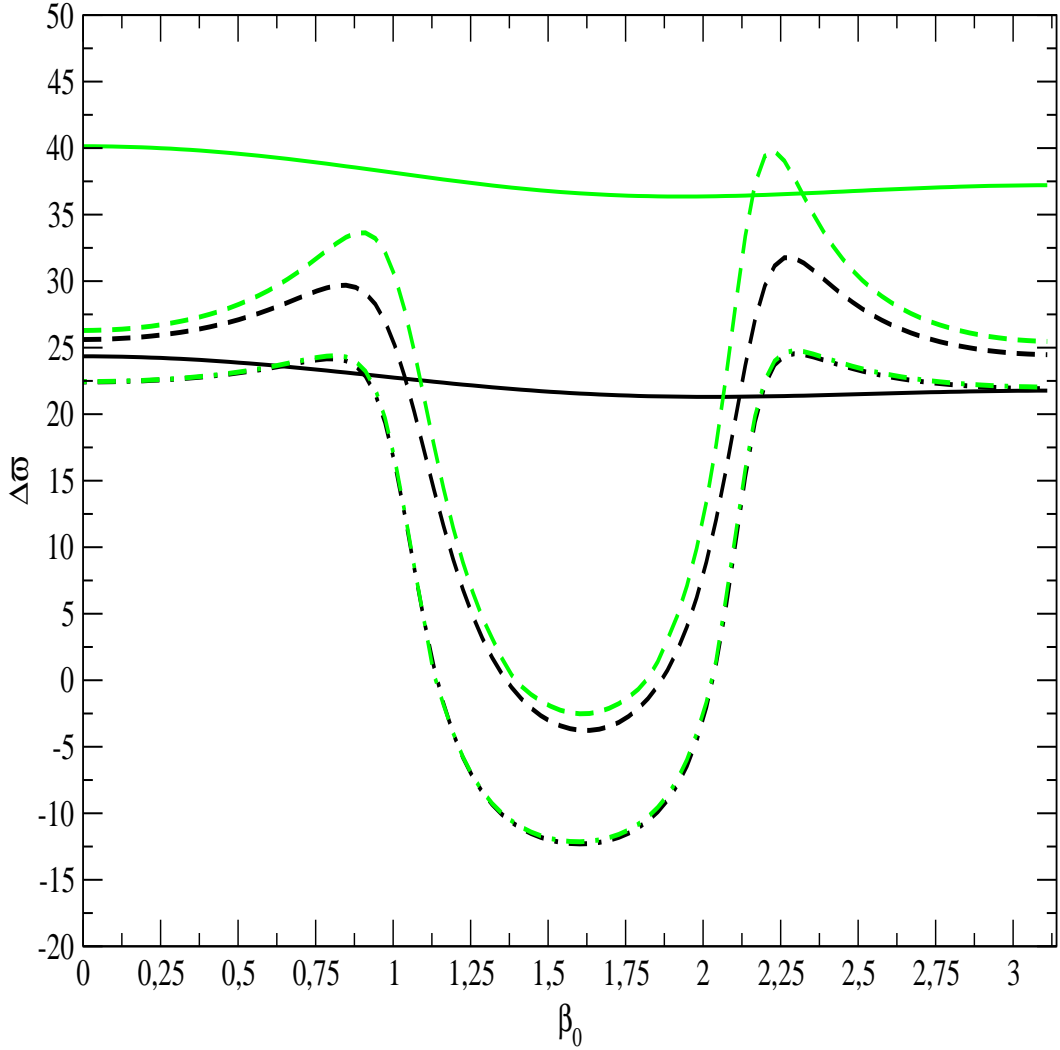}
\caption{ {As for} Fig. \ref{fig1}, but the final values of the difference $\Delta \varpi$ are shown.}
\label{fig2}
\end{center}
\end{figure}

\begin{figure}
\begin{center}
\vspace{1cm}
\includegraphics[width=14.0cm,height= 14.0cm,angle=0]{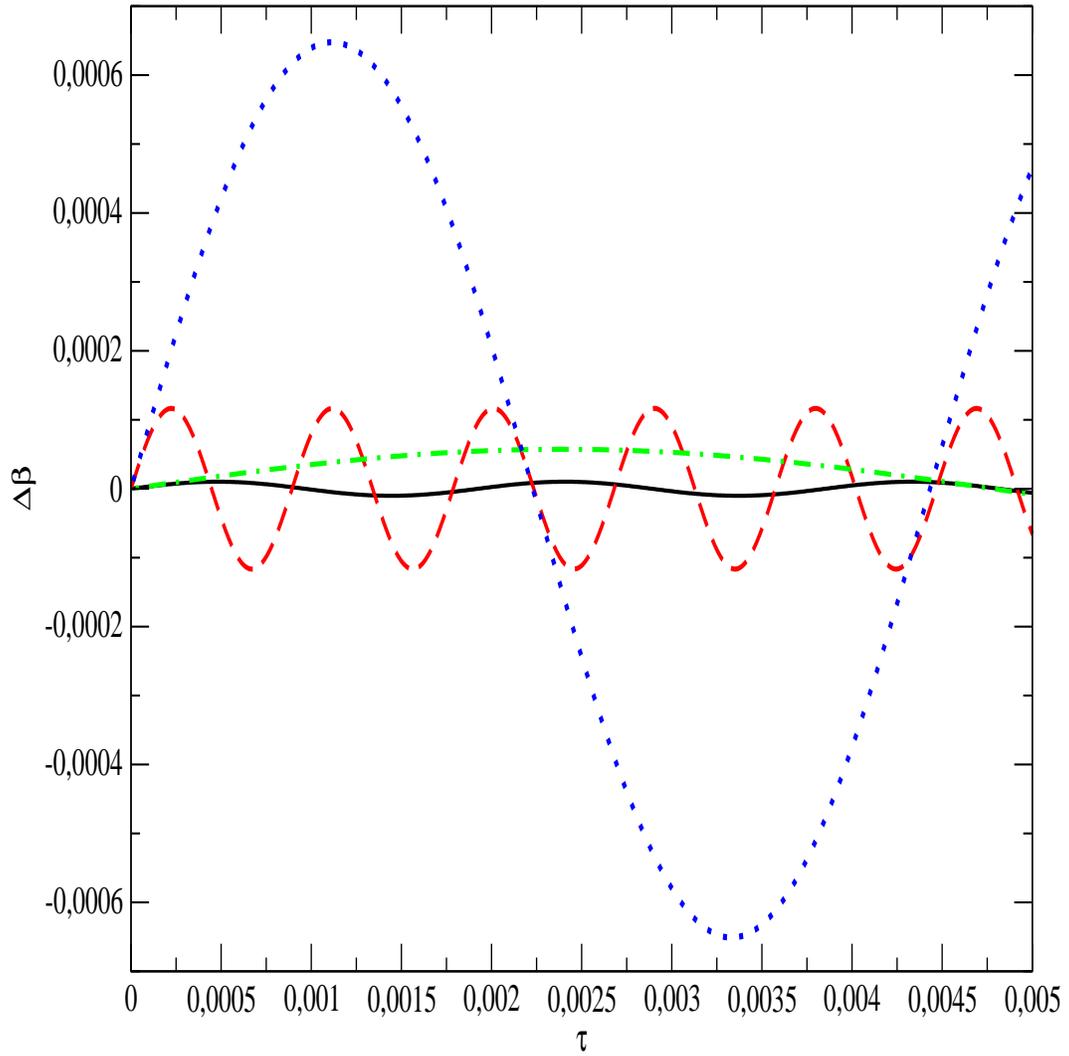}
\caption{This illustrates the time evolution of the difference $\Delta \beta=\beta-\beta_0$. 
All curves have $\tilde \Omega_r=3$. Solid and dashed curves have $\beta_0=0.5$ with $e_0=0.5$, $0.7$, respectively. Dot dashed and dotted curves 
have $\beta_0=1$ with $e_0=0.5$ and  $0.7$, respectively. Note that the latter value of $\beta_0$ is close 
to $\beta_{crit, +}$.}
\label{fig3}
\end{center}
\end{figure}

\begin{figure}
\begin{center}
\vspace{1cm}
\includegraphics[width=14.0cm,height= 14.0cm,angle=0]{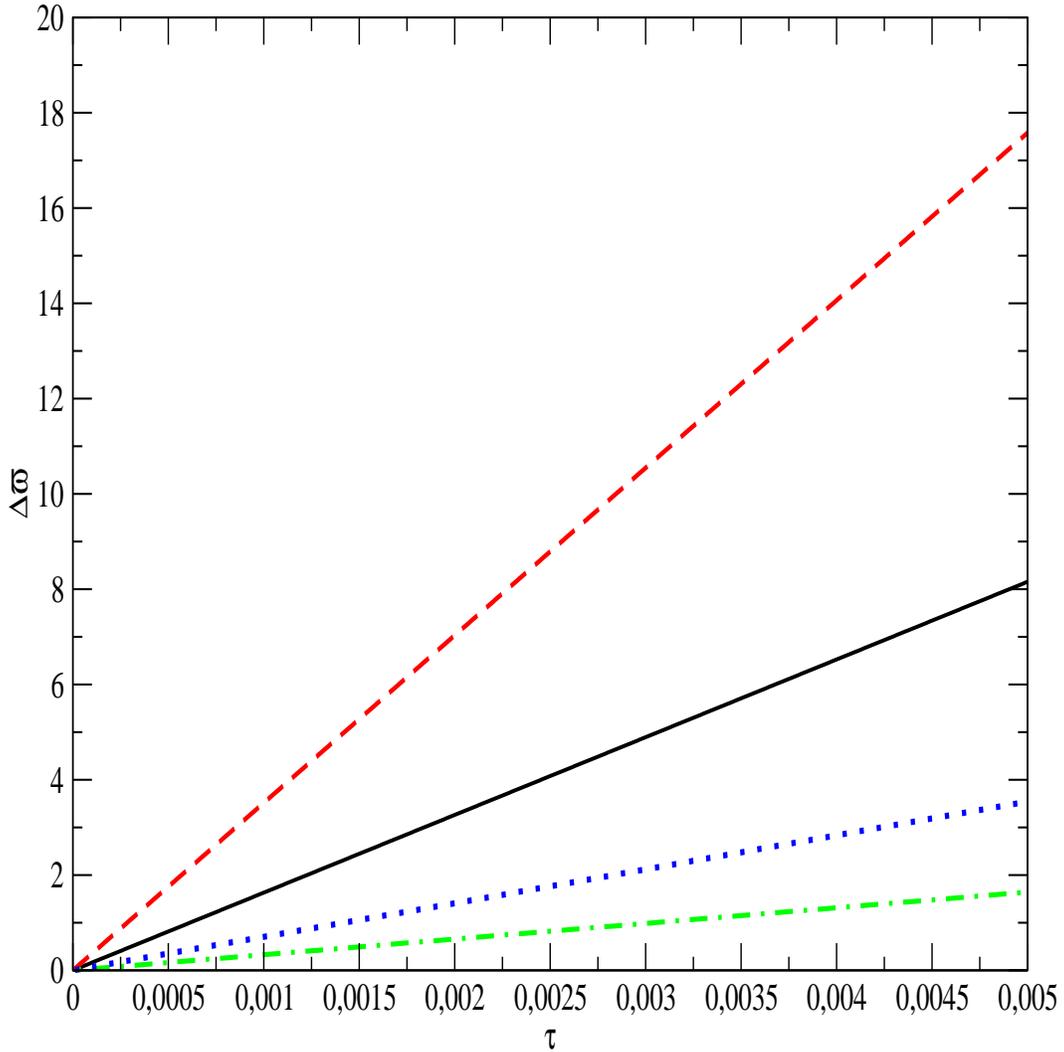}
\caption{The same as Fig. \ref{fig3}, but the relative values of the apsidal angle $\Delta \varpi=\hat \varpi -\varpi_0$ are 
shown.}
\label{fig4}
\end{center}
\end{figure}


We begin by considering  the case $q=10^{-3}$. In Figs. \ref{fig1} and \ref{fig2} we respectively show the maximum 
 value of $\Delta \beta =\beta-\beta_0,$.    $(\Delta \beta)_{max},$  over 
the time of computation and the final value of $\Delta \varpi=\varpi -\varpi_0$ 
as functions of $\beta_0$. 
One can see that $(\Delta \beta)_{max}$ is rather small in this case. This is especially the situation
for the small relative rotational frequency $\tilde \Omega_r=0.1$, where $(\Delta \beta)_{max} \sim 10^{-6}-10^{-5}$
depending on the value of the initial eccentricity. 
Curves corresponding to  $\tilde \Omega_r=1$ and $\tilde \Omega_r=3$ give similar
results with larger values of,   $(\Delta \beta)_{max},$ corresponding to larger initial  eccentricity. 
{The maximum value of the expression (\ref{dbet}) during a run} provides
a reasonable order of magnitude estimate of $(\Delta \beta)_{max}$ when $\beta_0 < \beta_{crit,+}$ and $\beta_0 > \beta_{crit,-}$.
In the intermediate range there are two maxima in the neighbourhood of $\beta_{crit,\pm}$ and a minimum in the neighbourhood of
 $\beta_0={\pi / 2}$. From Fig. \ref{fig2} one can see that the final apsidal angle increases when $\beta_0 < \beta_{crit,+}$ and $\beta_0 > \beta_{crit,-}$,
  decreases in the intermediate range and in close to zero when $\beta_0\sim \beta_{crit,\pm}$. 
  The latter observation
 { accounts for}  the existence of two peaks in the distribution of $(\Delta \beta)_{max}$ observed in Fig. \ref{fig1}.

In Fig. \ref{fig3} and \ref{fig4} we show the dependence of $\Delta \beta $ and $\Delta \varpi$ on time respectively.
The results are for  $\beta_{0}=0.5$ and $1$, and with  $e_0=0.5$ and $0.7$ in each case.
 One can see that $\Delta \beta$ is periodic with small amplitude 
in all cases, the largest amplitude of variation of $\Delta \beta$ is seen in the case of $\beta_0=1$ and $e=0.7$
consistently  with the results shown in the previous Figures. From Fig. \ref{fig4} it is seen that the dependence 
of $\hat \varpi$ on time is practically linear in all cases. The rate of change of $\hat \varpi$ increases with 
increase of initial eccentricity and decreases with increase of $\beta_0$ as expected.

\subsection{The case $q=1$}

\begin{figure}
\begin{center}
\vspace{1cm}
\includegraphics[width=14.0cm,height= 14.0cm,angle=0]{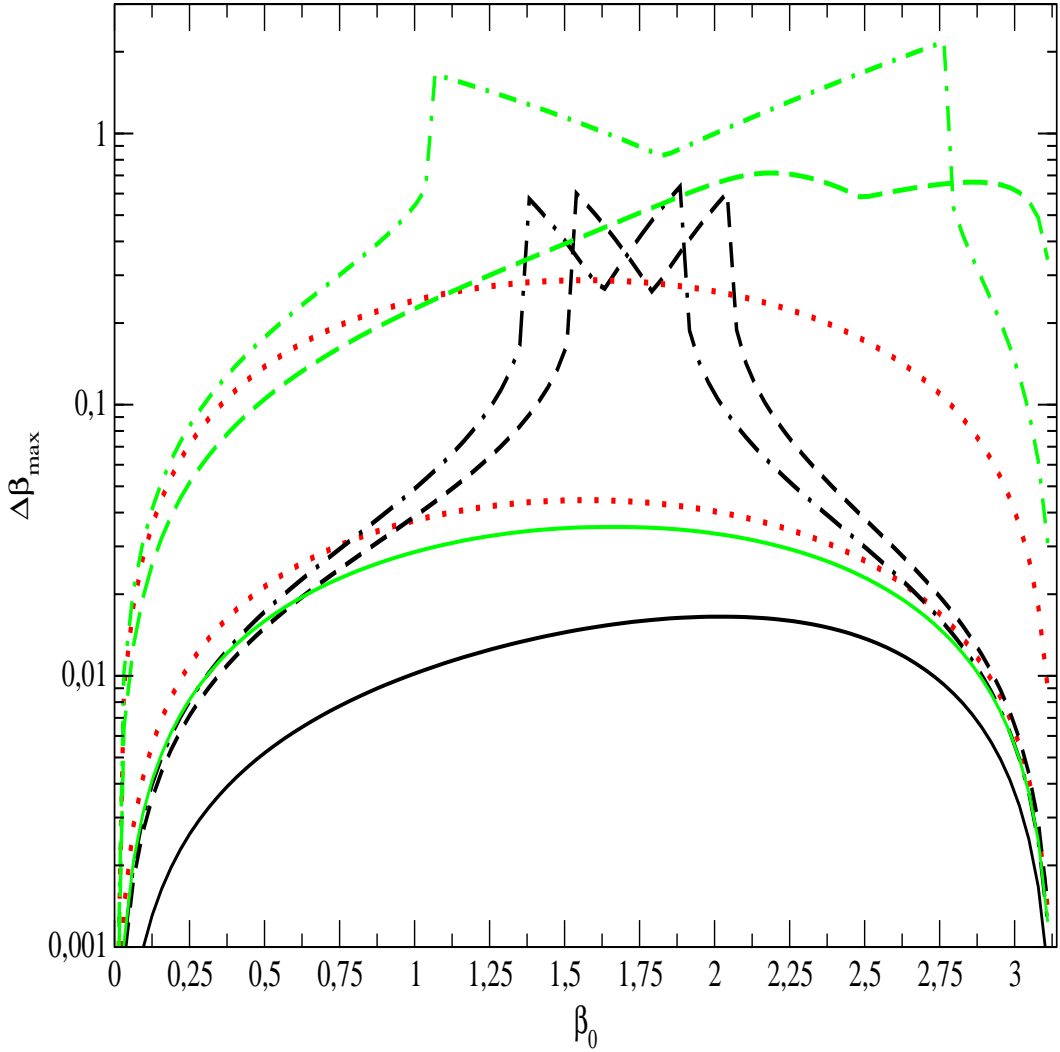}
\caption{Same as Fig. \ref{fig1}, but for $q=1$.}
\label{fig12}
\end{center}
\end{figure}

\begin{figure}
\begin{center}
\vspace{1cm}
\includegraphics[width=14.0cm,height= 14.0cm,angle=0]{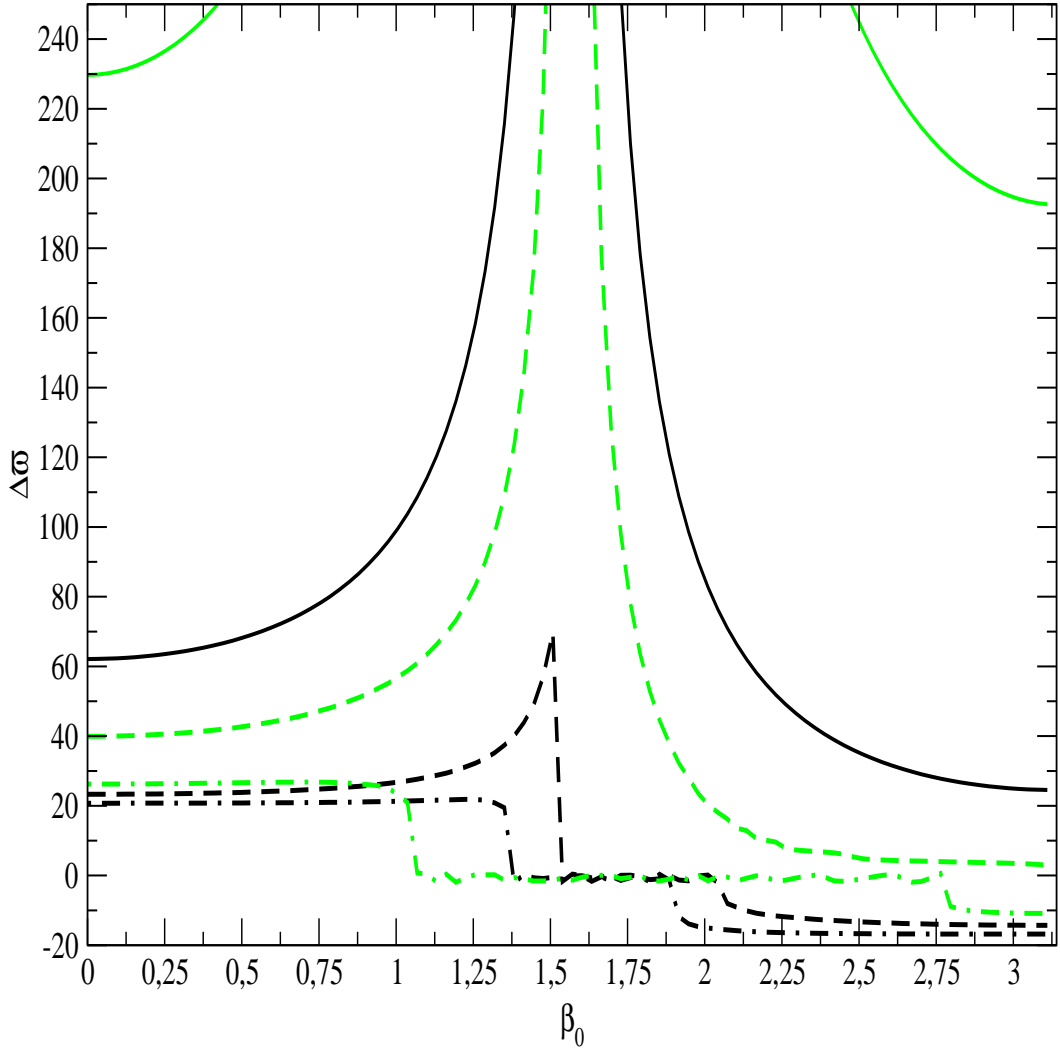}
\caption{Same as Fig. \ref{fig2}, but for  $q=1$.}
\label{fig22}
\end{center}
\end{figure}

\begin{figure}
\begin{center}
\vspace{1cm}
\includegraphics[width=14.0cm,height= 14.0cm,angle=0]{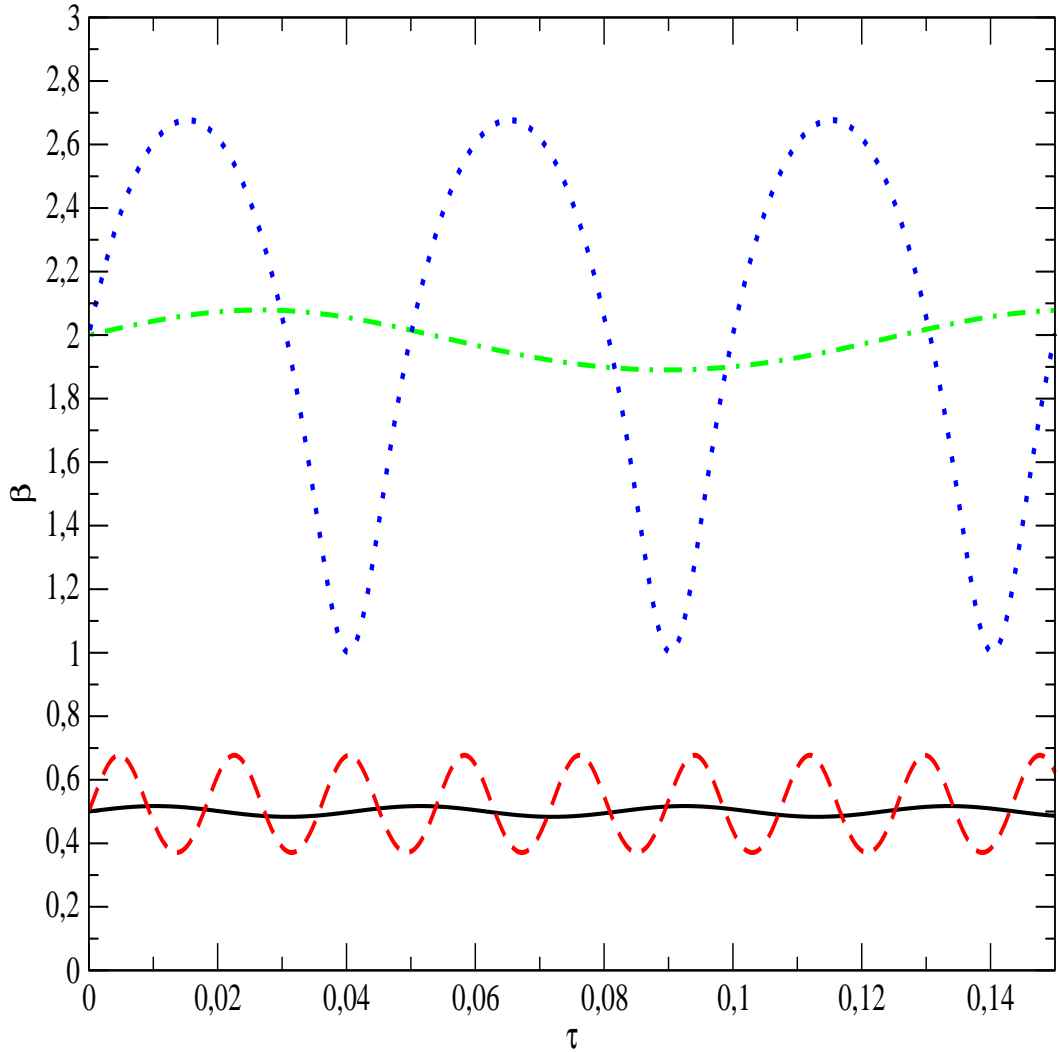}
\caption{As in  Fig. \ref{fig3}, but now we show the time dependence of the inclination angle
$\beta$ for $q=1$ with $\beta_0=0.5$ ( solid and dashed curves ) and $\beta_0=2$ (dot dashed and dotted curves).
{ As for Fig. \ref{fig3}  the larger amplitude variations are for the larger value of $e_0.$}}
\label{fig32}
\end{center}
\end{figure}

\begin{figure}
\begin{center}
\vspace{1cm}
\includegraphics[width=14.0cm,height= 14.0cm,angle=0]{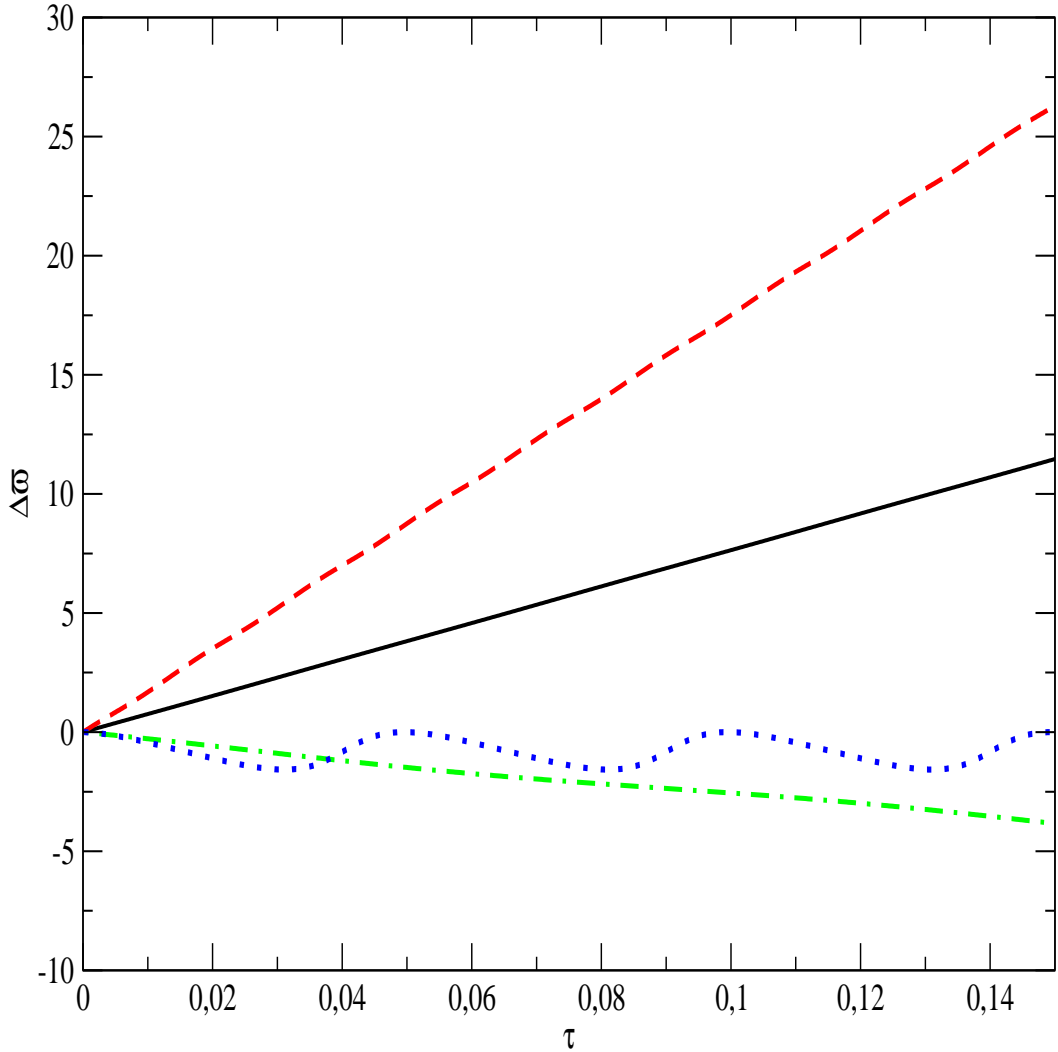}
\caption{As for  Fig. \ref{fig32}, but $\Delta \varpi$ is shown.}
\label{fig42}
\end{center}
\end{figure}

 In Figs. \ref{fig12} and \ref{fig22} we respectively  show the  {maximum value of}  $\Delta \beta =\beta-\beta_0$ 
and {the final value of}  $\Delta \varpi=\varpi -\varpi_0$ 
as functions of $\beta_0$ calculated for $q=1$. One can see that the situation is quite different from the previous
case. Although the curves corresponding to $\tilde \Omega_1=0.1$ indicate modest variation of $\beta$ on the  order of
$\Delta\beta \sim 10^{-2}$, when $\tilde \Omega_1=1$ or $3$ variations can be significantly larger when $\beta_0 > \beta_{crit,+}$. 
Only when $\beta_0 < \beta_{crit,+}$ {does the maximum value of the expression (\ref{dbet})}  give a correct order of magnitude estimate for the values of
$e_0$ considered. When $e_0=0.5$ it also gives a correct estimate when $\beta > \beta_{crit,-}$.
  Otherwise  variations 
are much larger. This effect is interpreted as the evolution of the solutions in {the regime  where  a critical
curve is crossed.}  
Let us recall that this is possible only when at some particular moment of time $\beta > \beta_{crit,+}$. When
this regime is realised we expect final values of $\Delta \varpi$ to be {limited and to remain} close to zero {in relative terms},
 since it is accompanied 
by libration of the apsidal angle. This is seen in Fig. \ref{fig22}, where we have $\Delta \varpi$ close to zero
for a range of $\beta_{0}$ in the interval $(\beta_{crit,+}, \beta_{crit,-})$ when $e_0=0.5$ 
and $\tilde \Omega_r=1$, or $3$.
 When $e_0=0.7$ we find  $\Delta \varpi$ close to zero when $\tilde \Omega_{r}=3$ and $\beta_{crit,+}<\beta <\sim
2.75$. The case $e_0=0.7$, $\tilde \Omega_r=1$ has the usual circulating regime of evolution of $\hat \varpi$, but 
the evolution rate  is tending asymptotically to zero when $\beta_0 >\sim 2$. This explains the large values of $(\Delta \beta)_{max}$ in this region of the parameters space.

We further illustrate the case $q=1$ in Figs \ref{fig32} and \ref{fig42}, where the evolution of $\beta $ and 
$\Delta \varpi$ are shown for $\beta_0=0.5$ (solid and dashed curves with $e=0.5$ and $0.7$, respectively) and
$\beta_0=2$ (dot dashed and dotted curves with  $e=0.5$ and $0.7$, respectively).  In all cases $\tilde \Omega_r=3$.
From the discussion above it follows that we expect  evolution with librating apsidal angle 
when $\beta_0=2$ and $e=0.7$.
 As seen from Fig. \ref{fig42} this conclusion is confirmed and Fig. \ref{fig32} 
shows that in this case variations of $\beta$ are especially large. Since this angle crosses ${\pi /2}$
in the course of evolution the system changes its state from prograde to retrograde rotation and vice verse. 
For an extensive analytic discussion of this regime see \cite{IP1}.

\subsection{The case $q=10^3$}    

\begin{figure}
\begin{center}
\vspace{1cm}
\includegraphics[width=14.0cm,height= 14.0cm,angle=0]{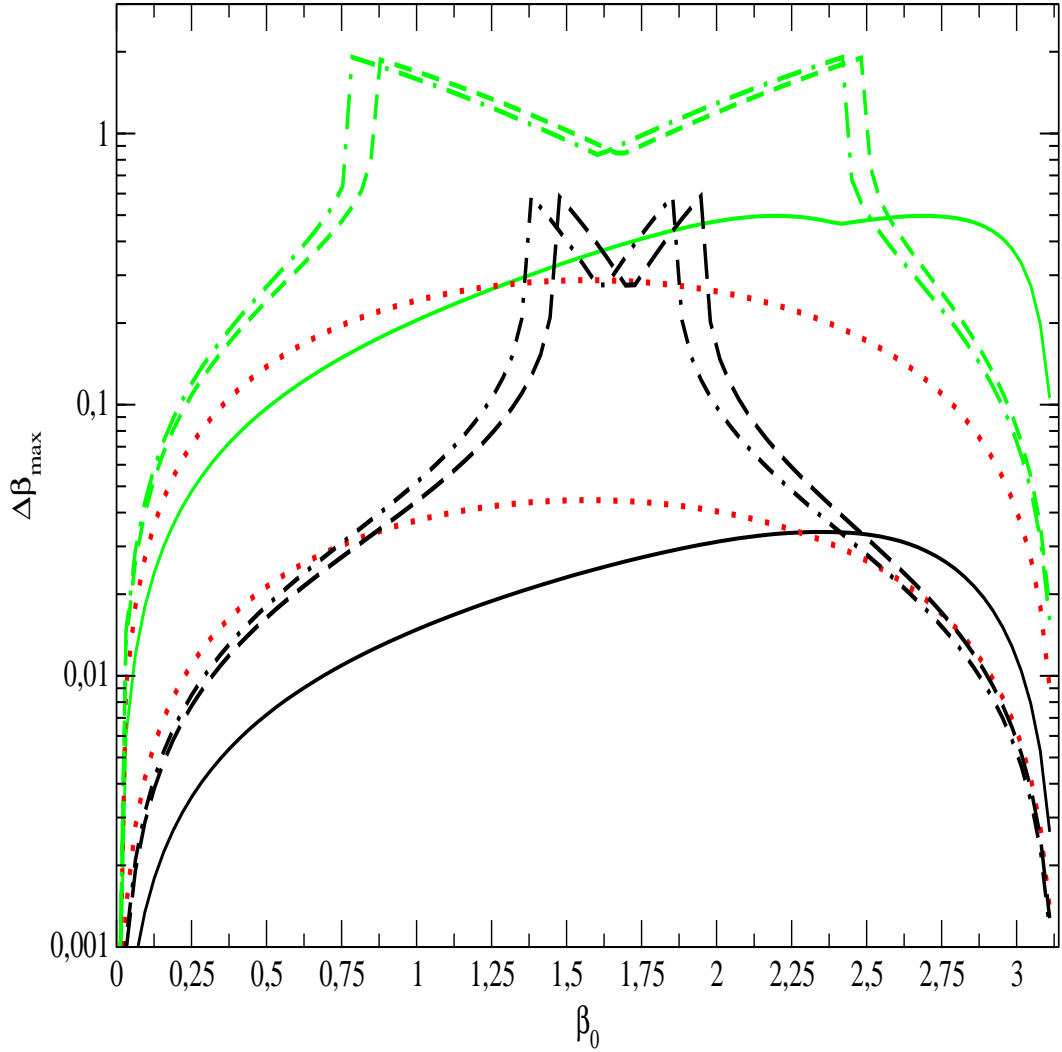}
\caption{As for  Figs. \ref{fig1} and \ref{fig12}, but $q=10^3$ and $\tilde a=50$.}
\label{fig13}
\end{center}
\end{figure}

\begin{figure}
\begin{center}
\vspace{1cm}
\includegraphics[width=14.0cm,height= 14.0cm,angle=0]{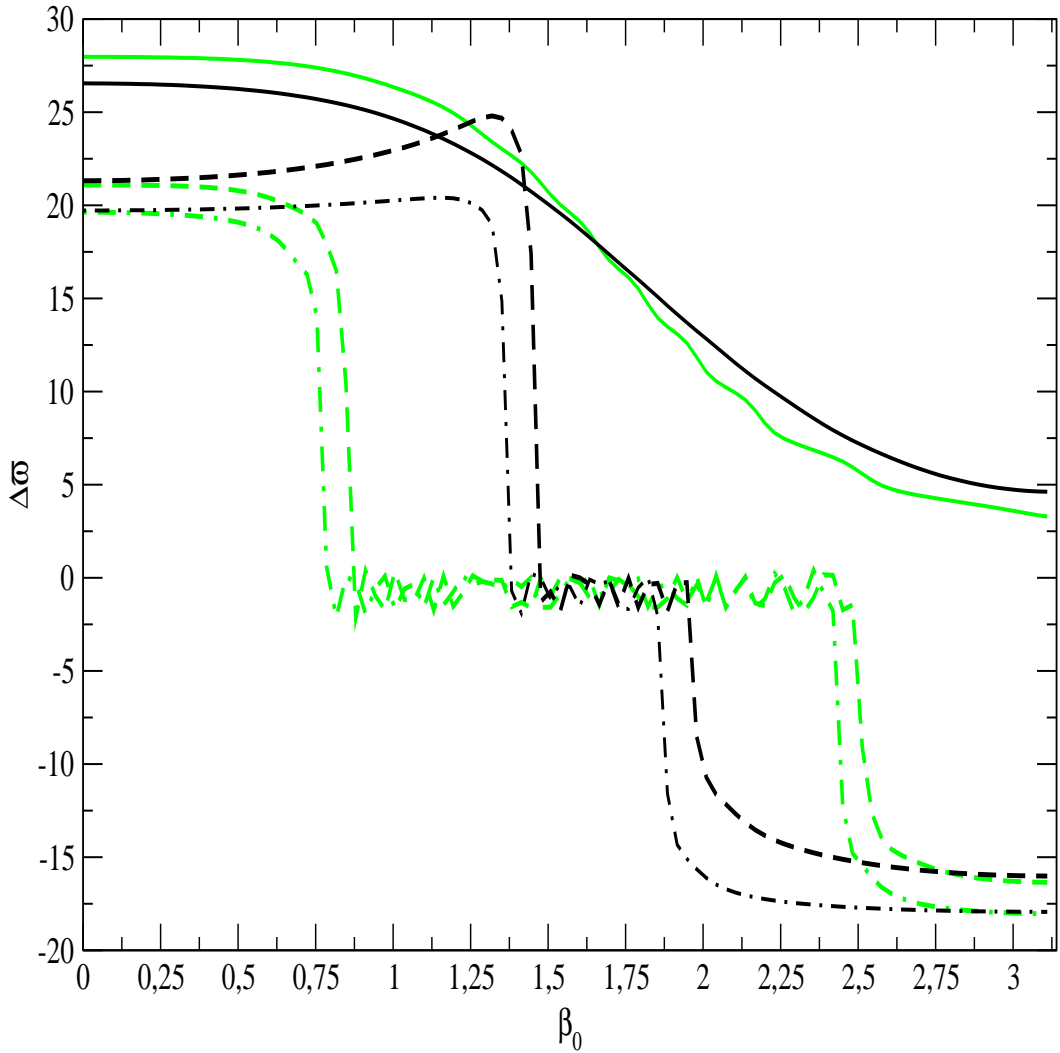}
\caption{As for Figs. \ref{fig2} and \ref{fig22}, but $q=10^3$ and $\tilde a=50$.}
\label{fig23}
\end{center}
\end{figure}

\begin{figure}
\begin{center}
\vspace{1cm}
\includegraphics[width=14.0cm,height= 14.0cm,angle=0]{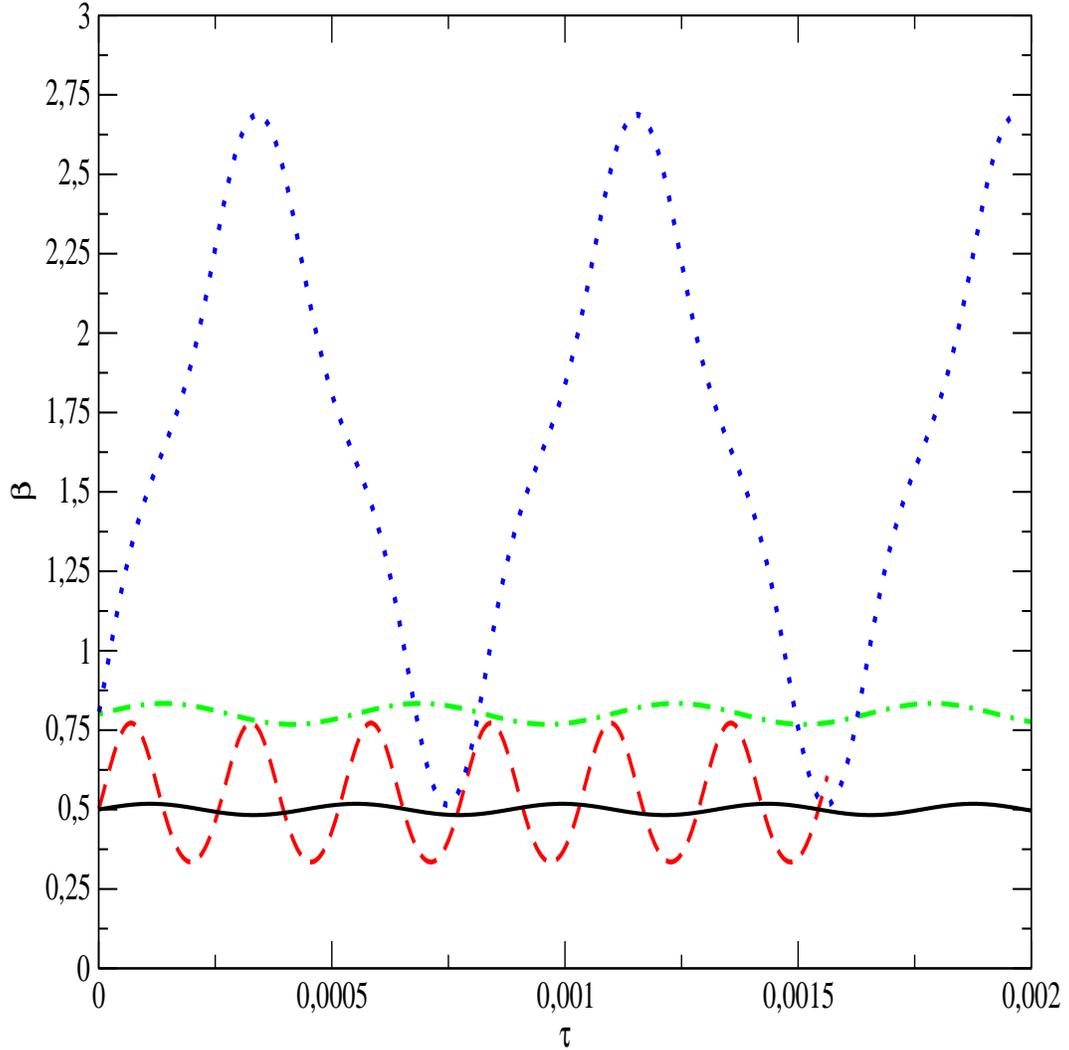}
\caption{As for  Fig. \ref{fig32}, but for $q=10^{3}$ with $\beta_0=0.5$ { ( solid and dashed curves) and $\beta_0= 0.8$  
(dot dashed and dotted curves)}.}
\label{fig33}
\end{center}
\end{figure}

\begin{figure}
\begin{center}
\vspace{1cm}
\includegraphics[width=14.0cm,height= 14.0cm,angle=0]{varpi_q+3.eps}
\caption{As for Fig. \ref{fig33}, but $\Delta \varpi$ is shown.}
\label{fig43}
\end{center}
\end{figure}

The general behaviour of $(\Delta \beta)_{max}$ and $\Delta \varpi$ in the case $q=10^{3}$ is shown in Figs. \ref{fig13} and
\ref{fig23}. Note that unlike the previous cases now $\tilde a$ is ten times large, $\tilde a=50$. Since in this
case it is assumed that non-dissipative tides are active in a planet of Jupiter size, this choice of $\tilde a$ 
corresponds to approximately the same physical value of semi-major axis as the other cases. In this case especially large variations
of $\beta$ are seen when $1.25 \sim < \beta_0 \sim < 2$ with  $e_0=0.5$ and when $0.75 \sim < \beta_0 \sim < 2.5$
with  $e_0 = 0.7$, see Fig. \ref{fig13}.  From Fig. \ref{fig23} it is seen that indeed in these ranges of $\beta_0$
$\Delta \varpi$ is close to zero, and, therefore, the regime of  apsidal angle libration is realised.

Figs \ref{fig33} and \ref{fig43} show the time dependence of $\beta$ and $\Delta \varpi$  for $\beta_0=0.5$ (solid and dashed curves have  $e_0=0.5$ and $0.7$, respectively) and
$\beta_0=0.8$ (dot dashed and dotted curves have  $e_0=0.5$ and $0.7$, respectively).
As in the previous case the regime
of  apsidal angle libration is expected  when $\beta_0=0.8$ and $e_0=0.7$. Fig. \ref{fig43} confirms this prediction,
while from Fig.  \ref{fig33} we see that variations of $\beta$ are quite large in this case and the system changes its 
rotation from prograde to retrograde, very much as in the case $q=1$.

\section{Discussion and Conclusions}   
\label{Discussion}
In this paper  we  reviewed and extended the main results of our paper \citep{IP}.
 { There a self-consistent approach to the evolution of the orbital parameters of an eccentric binary with misaligned orbital and spin angular momenta due to quasi-stationary tides was proposed. 
We described and reviewed  both  qualitatively  and quantitatively 
the new non-dissipative effects obtained within  the framework of our approach, giving
the full evolution equations  governing them 
 in Section \ref{Detorbspinev} .
These effects result from  the rotation of the primary.}
They  cause an evolution of the angle of  inclination between the orbital and spin angular momenta,
 $\beta,$ on a relatively short time scale
provided that the  source of apsidal precession is specified. 

Unlike the analysis made in \cite{IP} we took into account all potentially important sources of apsidal precession 
for an isolated binary with only one tidally active component.  {These are   Einstein precession, the apsidal 
precession rate  caused by the  tidal bulge, and that arising from rotational distortion, see equation (\ref{prec}).} Since the apsidal precession rate
due to stellar rotation depends  on the inclination angle, $\beta,$  in general the dynamical system describing 
the non-dissipative tidal evolution   {effectively has  two degrees of freedom. }
The presence of non-linear interaction
between two angles can lead to qualitatively new effects for certain values of the parameters of the system.

We illustrate these effects {by solving the corresponding dynamical equations numerically for an indicative exploratory set of input parameters
over a limited time span.}.
 We find that when mass ratio
$q$ is small variations of $\beta$ are expected to be small, and the apsidal angle changes monotonically with time. 
However, when the mass ratio is order of unity or larger and both stellar rotation and eccentricity are significant
periodic variations of $\beta$ can be order of unity, and the system can change the direction of its rotation from prograde to retrograde over some relatively short period of time. 
In the same situation the apsidal angle changes periodically 
with time. As explained in an accompanying paper \cite{IP1}, where we provide an extensive analytic analysis 
of the dynamical system,  these new effects are related to the possibility of the  existence of 'critical curves' - {the 
curves in  the parameter space of the system,} where the total apsidal precession rate \ref{prec} is zero.
 When the system crosses such a curve during its evolution the apsidal angle librates, while a typical change
of $\beta$ is much larger than that obtained from naive estimates.  
   Observational implications of these results will be reported elsewhere. 

There is a simple physical explanation for these new non-dissipative effects. Namely, in the absence of rotational 
effects (and, of course, while neglecting dissipation) the tidal bulge is
aligned with the direction to perturbing body. 
When the rotation axis is misaligned with respect to orbital angular momentum, {the presence of
rotational effects, arising from e.g. Coriolis forces,  produce torques
which lead to the evolution of the angle of inclination between the orbital and spin angular momenta.
These do not occur in the aligned case.} 
This resembles in part the well known Lidov-Kozai effect, but, in our case there is no need for the presence of a third body to break the symmetry of gravitational field.  

It is important to stress that we have neglected a contribution of toroidal displacements potentially excited by perturbation of the star
due to tides. Although it may be small  effect due to the  relatively  small magnitude 
of the appropriate overlap integrals, this contribution should be
separately analysed. A convenient framework for such an analysis 
would be the self-adjoint approach to the problem of tidal excitation of normal modes of any kind in a rigidly rotating star put forward in \cite{Papaloizou2005} and \cite{IP2007}.

{Finally, let us estimate typical values of $\Delta \beta$ and corresponding evolution times for  two  potentially interesting  systems.
 We  consider systems containing a neutron star with its stellar companion,  having a  rotation axis significantly misaligned with the orbital angular momentum, such that the inclination angle
  $\beta_{0} \sim 1$. In all cases we assume, for definiteness, that stellar rotation frequency is equal to the orbital angular frequency, and accordingly $\tilde \Omega_r=1$. 
   The evolution time scale of $\Delta \beta$, $t_{ev}$, is here defined as $t_{ev}={\rm \pi}| ({d\varpi / dt})^{-1}|$, where ${d\varpi / dt}$ is given by eq. (\ref{prec}) and it is also assumed that a single term on r.h.s of (\ref{prec}) dominates. To estimate characteristic values of $\dot \varpi_{R}$ and $\dot \varpi_{NR}$ we set $3\cos^2 \beta -1$ and $\cos \beta$ to be unity in the corresponding expressions, see eq. (\ref{a15}).

At first we consider a system with parameters similar to the parameters of the famous low mass X-ray binary
Sco X-1. At the present time its eccentricity $e$ is smaller than $\sim 10^{-2}$, see e.g. \cite{MN} and references therein. However, it is reasonable to assume that it has been significantly larger in the past, due to a possible kick during the supernova explosion leading to the formation of the neutron star. The same kick could be responsible for the spin misalignment. 
To take into account the system's evolution we use initial masses of the donor star, its radius, mass of the neutron star and the orbital period from Table 1 of \cite{FT}, their model 4, which is realistic according to these authors. Namely, we use $M_*=0.7M_{\odot}$, $R_{*}=0.53R_{\odot}$,  $M_p = 1.41M_{\odot}$ and $P_{orb}=10h$. 
The values adopted for $k_2$ and ${\tilde I}$ in this Section are  $0.01$ and $0.1$ respectively.   

We consider two values of eccentricity, $e=0.2$ and $e=0.5$, and find that in both cases  ${d\varpi / dt}$ is dominated by the non-inertial term, $\dot \varpi_{NI}$ and use, accordingly, eq. (\ref{dbet})  to estimate $\Delta \beta$. We obtain  $\Delta \beta \approx 4.5\cdot 10^{-3}$ and $t_{ev}\approx 0.49yr$ for $e=0.2$,  and  $\Delta \beta \approx 6.1 \cdot 10^{-2}$ and $t_{ev}\approx 0.34yr$ for $e=0.5$. it is clear that $t_{ev}$ is much smaller than any characteristic time scale of secular orbital
evolution driven by tidal effects. Note that other terms in (\ref{prec}) are smaller than the leading one by at least one order of 
magnitude for such a system. In this case the presence of the critical curves leading to larger values of
$\Delta \beta $ is not expected unless $\beta$ is extremely  close to ${\rm \pi}/2.$

Next we consider the system GX-301-2, which consists of a hypergiant companion and an X-ray pulsar. We
use $M=43M_{\odot}$, $R_{*}=68R_{\odot}$,  $M_p=1.85M_{\odot}$ and $P_{orb}=45d$ as well as $e=0.5$, see e.g. \cite{GX1}, \cite {GX2} and \cite{GX3}. We see again that  ${d\varpi / dt}$
is dominated by non-inertial term, but, in this case the next to  leading term, $\dot \varpi_{T,}$ is just a factor of two  smaller, thus there is a possibility of being near  a  critical curve. Nonetheless, we use (\ref{prec}), (\ref{a15}) and (\ref{dbet}) to estimate $t_{ev}$ and $\Delta \beta$ noting that for some plausible system  parameters and  significant inclination, $\beta,$  $\Delta \beta$ obtained from (\ref{dbet}) could be significantly underestimated. We obtain $t_{ev}\approx 150 yr$ and $\Delta \beta \approx 1.1\cdot 10^{-2}$. It is worth stressing again that $t_{ev}$ is likely to be much smaller than any potential time scale of secular orbital evolution. We note that the hypergiant star in this system exhibits a significant mass loss due to stellar wind, with corresponding mass loss rate $ \dot M_{W}\sim 10^{-5}M_{\odot}/yr$, see e.g. \cite{GX2}. This is expected to cause orbital evolution on time scale $\sim M_*/\dot M_{W}\sim 10^{5}-10^{6}yr$. This is two-three orders of magnitude larger than $t_{ev}$.}            

\section*{Acknowledgments}

PBI was supported in part by the grant 075-15-2020-780 'Theoretical
and experimental studies of the formation and evolution
of extrasolar planetary systems and characteristics of exoplanets'
of the Ministry of Science and Higher Education of the Russian Federation.
 We are grateful to A. J. Barker, K. A. Postnov, N. I. Shakura and { the referee} for useful comments.






\label{lastpage}

\end{document}